

\overfullrule=0in
\magnification=\magstep1
\centerline{\bf RANDOM WALKS IN NONINTEGER DIMENSION}
\bigskip
\bigskip
\bigskip
\centerline{Carl M. Bender and Stefan Boettcher\footnote{$^*$}{Current
Address: Brookhaven National Laboratory; Upton, NY 11973}}
\medskip
\centerline{Department of Physics}
\medskip
\centerline{Washington University}
\medskip
\centerline{St. Louis, MO 63130-4899}
\bigskip
\centerline{Lawrence R. Mead}
\medskip
\centerline{Department of Physics and Astronomy}
\medskip
\centerline{University of Southern Mississippi}
\medskip
\centerline{Hattiesburg, MS 39406-5046}
\bigskip
\bigskip
\bigskip
\bigskip
\bigskip
\centerline{\bf ABSTRACT}
\bigskip
One can define a random walk on a hypercubic lattice in a space of integer
dimension $D$. For such a process formulas can be derived that express the
probability of certain events, such as the chance of returning to the origin
after a given number of time steps. These formulas are physically meaningful
for
integer values of $D$. However, these formulas are unacceptable as
probabilities
when continued to noninteger $D$ because they give values that can be greater
than $1$ or less than $0$. In this paper we propose a random walk which gives
acceptable probabilities for all real values of $D$. This $D$-dimensional
random
walk is defined on a rotationally-symmetric geometry consisting of concentric
spheres. We give the exact result for the probability of returning to the
origin
for all values of $D$ in terms of the Riemann zeta function. This result has a
number-theoretic interpretation.
\footnote{}{PACS numbers: 05.40.+j, 03.29+i, 02.50+5}
\footnote{}{hep-lat/9311011}
\vfill \eject

\noindent {\bf I. INTRODUCTION}
\bigskip

Recent studies of quantum field theory in $D$-dimensional Euclidean space have
focussed on the possibility of expanding the Green's functions of the theory as
series in powers of $D$.$^{1,2,3}$ A technique used in this work is to
introduce
a $D$-dimensional hypercubic lattice, and, using strong-coupling lattice
methods, to obtain expressions for the Green's functions for integer values of
$D$. Order by order in these strong-coupling expansions the coefficients are
polynomials in $D$ and therefore appear to have simple and natural
continuations
to noninteger values of $D$. Of course, such a procedure is not unique; there
may be many different analytic continuations off the integer-dimensional
hypercubic
lattices. The procedure was justified on the basis of comparisons with known
results in solvable quantum field theoretic models.

Such an analytic continuation does not always produce acceptable physical
results. For example, if we try to define a random walk in
arbitrary-dimensional
space by analytically continuing a random walk off an integer-dimensional
hypercubic lattice, we can obtain unphysical values for probabilities for
certain physical processes. To illustrate very simply how such problems can
arise, consider the probability $P_2$ that a random walker on a lattice in
$D$-dimensional space will return to the starting point after two steps. To
compute this probability we consider $D$ to be a positive integer and the
lattice to be hypercubic. For such a lattice,
$$P_2 = {1\over {2D}}~~. \eqno\hbox{(1.1)}$$
The obvious analytic continuation of this result to noninteger $D$, namely
the function $1/(2D)$, cannot be the correct probability when $D<1/2$.

The purpose of this paper is to devise a $D$-dimensional geometry in which a
random walk is well defined when $D$ is noninteger. In this model all
probabilities have sensible values; they lie between $0$ and $1$.$^4$

This paper is organized as follows: In Sec.~II we review the conventional
treatment of random walks on integer-dimensional hypercubic lattices. We show
that if we
attempt to continue the results to noninteger dimension we obtain unacceptable
values of probabilities for physical events. In Sec.~III we invent a
spherically
symmetric geometry on which a random walk is well defined for all real
values of $D$. This geometry eliminates the need for analytic continuation. We
show how to express the probabilities in this model as solutions to a
difference
equation. We solve this equation and show that for all real $D$, probabilities
have physically acceptable values. One quantity of particular interest is
$\Pi_D$, the probability that a random walker will eventually return to the
origin. We determine $\Pi_D$ in closed form for all values of $D$. For one
interesting special case, in which the random walk occurs on a geometry of
concentric spheres whose radii are consecutive integers,
$$\Pi_D = 1 - {1\over {\zeta (D-1)}} \quad (D\geq 2)~~,\eqno\hbox{(1.2)}$$
where $\zeta$ is the Riemann Zeta function.
This result has an interesting number-theoretic interpretation.
\vfill \eject
\noindent {\bf II. RANDOM WALKS ON HYPERCUBIC LATTICES}
\bigskip

In this section we review the standard formalism$^5$ for describing random
walks
on a $D$-dimensional hypercubic lattice, where $D$ is an integer. Let
$C_{{\bf n},t}$ represent the
probability of the random walker being at the position ${\bf n}$ at time $t$.
Note that ${\bf n}$ is a $D$-component vector having integer entries and
that $t$ is a nonnegative integer. The random walk is assumed to begin at
the origin at $t=0$ so that
$$C_{{\bf n},0} = \delta_{{\bf n},{\bf 0}}~~.\eqno(2.1)$$
The random walk consists of unit-length steps connecting adjacent vertices
on the lattice. Thus, at each lattice point the walker may go in any one of
$2D$ directions with equal probability.

To describe a random walk we construct a recursion relation satisfied by the
probabilities $C_{{\bf n},t}$:
$$C_{{\bf n},t} = {1\over {2D}} \sum_{j=1}^D \left ( C_{{\bf n}-{\bf
e}_j,t-1} + C_{{\bf n} + {\bf e}_j,t-1} \right )~~,\eqno(2.2)$$
where ${\bf e}_j$ is the $D$-component unit vector for which all entries are
zero except for the $j$th component which is 1.
This equation states that the chance of being at the position ${\bf n}$ at
time $t$ is the sum of the probabilities of having been at $2D$
equally-likely neighboring sites.

To solve this recursion relation we introduce the discrete Fourier transform;
to
wit,
$$C_{{\bf n},t} = \int_{-\pi}^{\pi} {{d^Dk}\over {(2\pi)^D}} {\cal
X}_t ({\bf k})
e^{-i{\bf k}\cdot {\bf n}}~~,\eqno(2.3)$$
where ${\bf k}$ is now a continuous $D$-dimensional vector. In terms of
${\cal X}_t ({\bf k})$ the recursion relation reads
$${\cal X}_t ({\bf k})= {\cal X}_{t-1}({\bf k}) {1\over D}\sum_{j=1}^D \cos
k_j~~.\eqno(2.4)$$
The general solution to this equation is
$${\cal X}_t ({\bf k}) = A({\bf k}) \left ( {1\over D} \sum_{j=1}^D \cos
k_j \right ) ^{\displaystyle t}.\eqno(2.5)$$
To determine the constant $A({\bf k})$ we use the initial condition in
(2.1) which specifies that $A({\bf k})\equiv 1$.
We now obtain $C_{{\bf n},t}$ by taking the inverse Fourier transform
$$\eqalign{
C_{{\bf n},t}~&=~\int_{-\pi}^{\pi} {d^Dk \over (2\pi)^D}
\left ( {1 \over D} \sum_{j=1}^D \cos k_j \right )^{\displaystyle t}
e^{-i{\bf k}\cdot{\bf n}}~~\cr
&=~(\partial_x)^t \int_{-\pi}^{\pi} {d^Dk \over (2\pi)^D}
\exp\left ( -i{\bf k}\cdot{\bf n}+
{x \over D} \sum_{j=1}^D \cos k_j \right ) \Bigm \vert_{x=0}\cr
&=~\partial_x^t \prod_{j=1}^D I_{n_j}\left ({x \over D} \right
)\Bigm \vert_{x=0}~~,}\eqno(2.6)$$
where $I_{\nu}(x)$ is the modified Bessel function.

We are particularly interested in the probability of returning to the
origin ${\bf n}={\bf 0}$:
$$\eqalign{
C_{{\bf 0},2t}~&=~
\int_{-\pi}^{\pi} {{d^Dk}\over{(2\pi)^D}}~
\left ( {1\over D} \sum_{j=1}^D \cos k_j \right) ^{\displaystyle 2t}\cr
&=~\partial_x^{2t} \left [ I_0\left ({x \over D} \right ) \right ] ^D
\Bigm \vert_{x=0}~~, } \eqno(2.7)$$
where
$$I_0(x)=\sum_{l=0}^{\infty} {{(x/2)^{2l}}\over {(l!)^2}}~~.$$
(Note that $C_{{\bf 0},2t+1} = 0$.)
For $D=1$ for instance, we obtain the familiar result
$$C_{{\bf 0},2t}~=~ 2^{-2t} {{(2t)!}\over {(t!)^2}}~~,\eqno(2.8)$$
for the probability of a one-dimensional random walker to return to the origin
after $t$ time steps.

For arbitrary integer $D$ we have
$$\eqalign{
C_{{\bf 0},2}~&=~{1 \over 2D}~~,\cr
C_{{\bf 0},4}~&=~{1 \over {8D^3}} \left ( 6D-3 \right )~~,\cr
C_{{\bf 0},6}~&=~{5 \over {16D^5}} \left ( 6D^2-9D+4\right )~~.}\eqno(2.9)$$
Observe that for some noninteger values of $D$ these expressions are not
acceptable probabilities. Specifically, $C_{{\bf 0},2t}$ is oscillatory
until $D$ is sufficiently large, at which point it is positive and monotone
decreasing. On Fig.~1 we have plotted $C_{{\bf 0},2}$, $C_{{\bf 0},4}$,
and $C_{{\bf 0},6}$. These values become acceptable as probabilities only when
$D>1/2$, $D>1/2$, and $D> 0.7224\ldots$, respectively. In general, as $t$
increases, the lowest value of $D$ at which $C_{{\bf 0},2t}$ becomes an
acceptable probability also increases. On Fig.~2 we plot scaled versions of
$C_{{\bf 0},18}$ and $C_{{\bf 0},20}$ as functions of $D$. Note that
$C_{{\bf 0},18}$ becomes an acceptable probability only when $D$ is roughly
$2$ or larger, and $C_{{\bf 0},20}$ only when $D$ is roughly $2.6$ or larger.
A more precise picture of the behavior of $C_{{\bf 0},2t}$ is given in Table
1. In this table we give the critical values of the dimension $D_{\rm
critical}$
for which the function $C_{{\bf 0},2t}$ can first be considered a probability.
That is, for all $D>D_{\rm critical}$, $0 \leq C_{{\bf 0},2t} \leq 1$.

Having constructed the functions $C_{{\bf 0},2t}$ we can now construct
expressions $P_{2t}$ which represent the probability that the random walker
returns to the origin at time $2t$ for the {\sl first} time. The first few
such expressions are

$$P_2 = C_{{\bf 0},2}~~,$$
$$P_4 = C_{{\bf 0},4}- C_{{\bf 0},2}^2~~,$$
$$P_6 = C_{{\bf 0},6} - 2 C_{{\bf 0},2} C_{{\bf 0},4} + C_{{\bf 0},2}^3~~,$$
$$P_8 = C_{{\bf 0},8} - 2 C_{{\bf 0},2} C_{{\bf 0},6} - C_{{\bf 0},4}^2 +
3 C_{{\bf 0},2}^2 C_{{\bf 0},4} - C_{{\bf 0},2}^4~~.$$

In general the relation between $P_{2t}$ and $C_{{\bf 0},2t}$ is given by
the elegant formula$^5$
$$\sum_{t=1}^{\infty} P_{2t} x^t = 1 - {1\over {\sum_{t=0}^{\infty}
C_{{\bf 0},2t} x^t }}~~.\eqno{(2.10)}$$
Also, if we evaluate the above generating functions at $x=1$ we obtain
the probability $\Pi_D$ that the random walker ever returns to the origin:
$$\Pi_D = \sum_{t=1}^{\infty} P_{2t} = 1 - {1\over {\sum_{t=0}^{\infty}
C_{{\bf 0},2t}}}~~.\eqno{(2.11)}$$

The quantity $\Pi_D$ in (2.11) represents a physically interesting probability,
namely the chance that a random walker will eventually intersect her starting
point. However, as we can see, its mathematical form is not very satisfying
because it is presented in terms of a sum over probabilities $C_{{\bf 0},2k}$,
each of which is an unacceptable probability for some range of noninteger $D$.
Nevertheless, we can perform a formal Borel-like {\sl summation} of the series
in (2.11) by substituting the integral representation for $C_{{\bf 0},2t}$ in
(2.7) and interchanging the orders of summation and integration. We obtain
immediately a closed-form quadrature expression for $\Pi_D$:
$$\Pi_D = 1 - {1\over {\int_{0}^{\infty} dt\; e^{-t} [I_0(t/D)]^D }}~~.
\eqno{(2.12)}$$

We emphasize that the expression in (2.12) for $\Pi_D$ derives from an infinite
sum over probabilities that are unacceptable. However, when recast in the form
in (2.12) we find the surprising result that $\Pi_D$ is itself an acceptable
probability for all real positive $D$ if we interpret the expression as
follows. For $0 \leq D \leq 2$ it is clear that the integral in (2.12)
diverges. Thus, we conclude that
$$\Pi_D = 1 \quad (0\leq D\leq 2)~~.\eqno{(2.13)}$$
This conclusion verifies the well-known result, sometimes referred to as
Polya's
theorem, that a random walker in one or two dimensions will ultimately return
to
the origin with a probability of $1$.

For $D>2$ the integral in (2.12) exists and is a smooth function of $D$. We can
thus regard (2.12) as a {\sl definition} of $\Pi_D$ for noninteger $D$. We find
that for $D>2$, $\Pi_D$ is a smooth, monotone-decreasing function of $D$ and is
thus an acceptable probability function. For example, for the special case
$D=3$
we obtain$^5$
$$\Pi_3 = 1 - {{32 \pi^3}\over {\sqrt{6}\Gamma(1/24)\Gamma(5/24)\Gamma(7/24)
\Gamma(11/24)}} = 0.3405373296\ldots ~~.\eqno{(2.14)}$$

Unfortunately there is no way to evaluate the integral in (2.12) for arbitrary
$D$ or even arbitrary integer $D$. Thus, we are forced to study this integral
representation asymptotically. For large $D$, $\Pi_D$ has the asymptotic form
$$\Pi_D \sim {1\over{2D}} + {1\over{2D^2}} + {7\over{8D^3}} +
{{35}\over{16D^4}} + {{215}\over{32D^5}} + {{1501}\over{64D^6}} +
{{5677}\over{64D^7}} + {{44489}\over{128D^8}} + \ldots\quad (D\to\infty)~~.
\eqno{(2.15)}$$
This expansion is not accurate when $D$ is as small as $3$. However, it rapidly
becomes accurate as $D$ increases. The structure of (2.15) is not as simple
as it first appears; eventually the signs of the terms become negative, remain
negative for many orders, become positive for many orders, and appear to
continue oscillating in this block pattern.

Note that from the above description, the function $\Pi_D$ is also not smooth
at $D=2$; it has a discontinuous derivative there. When $D<2$, $\Pi_D=1$. On
the
other hand, when $D\geq 2$
$$\Pi_D \sim 1-{{\pi}\over 2}(D-2)+\ldots \quad (D\to 2^+)~~.$$
Thus, the function in (2.12) approaches its value at $D=2$ from above with the
negative
slope $-\pi/2$.

If we analytically continue the function in (2.12) from values of $D$ above $2$
to values below $2$ we obtain an unacceptable probability; we find that $\Pi_D$
rises as $D$ decreases and eventually becomes infinite at $D=1$ (see Appendix).
All of the above properties of the function $\Pi_D$ are summarized in Fig.~3.

Our objective in this paper is to construct a lattice geometry for which the
function $\Pi_D$ is an infinite sum of terms, all of which are {\sl acceptable}
probabilities for all real $D$ and not just integer $D$. Our construction is
given in the next section.
\vfill\eject
\noindent {\bf III. RANDOM WALKS IN ${\bf D}$ DIMENSIONS}
\bigskip

To define a random walk in $D$ dimensions, where $D$ is not necessarily an
integer, we return to the discussion of a random walk on a hypercubic lattice.
Normally, we regard a random walker as going from one lattice point on this
lattice to an adjacent point. However, for a completely equivalent way of
viewing the walk, consider the {\sl dual} lattice in which each lattice point
is
replaced by a $D$-dimensional hypercube of side $1$. The walker begins in the
center cube; we define $region~1$ as the interior of this cube. There are $2D$
cubes contiguous to $region~1$ into which the walker may go on the first step.
The region inside these $2D$ cubes is labeled $2$. On the second step the
walker
may return to $region~1$ or she may enter $region~3$, the region composed of
all
cubes contiguous to $region~2$ except for $region~1$. $Region~3$ comprises
$2D^2$ cubes. On the third step the walker may return to $region~2$ or she may
enter $region~4$, the region composed of all cubes adjacent to the cubes in
$region~3$ except for those cubes in $region~2$. There are $2D(2D^2+1)/3$ cubes
in $region~4$. In Fig.~4 we depict the first five of these regions for the case
$D=2$. Note that if the walker is in $region~n$, she must enter either
$region~n+1$ or
$region~n-1$ on her next step.  At each time step the walker measures
the area of the surfaces bounding the hypercube she occupies and travels
outward or
inward with a probability in exact proportion to the outward or inward surface
area she must traverse.

It is easy to visualize this model of a random walk in any geometry. The
simplest such geometry is, of course, a set of regions bounded by
concentric nested spheres. As we will see, such a geometry does not single out
the integer dimensions as being special in any way; a random walk can be
defined
in any real positive dimension $D$.$^6$

Specifically, we consider an infinite set of concentric nested spheres of
radius
$R_n$ ($n=1,~2,~3,~\ldots$). Then $region~n$ is the volume between the $n-1$st
and the $n$th spheres (see Fig.~5). Thus, if the walker is in $region~n$ the
probability of walking outward, $P_{\rm out}(n)$, and the probability of
walking
inward, $P_{\rm in}(n)$, are given by
$$\eqalign{P_{\rm out}(n)&={{S_D(R_n)}\over {S_D(R_n)+S_D(R_{n-1})}}~~,\cr
P_{\rm in}(n)&={{S_D(R_{n-1})}\over {S_D(R_n)+S_D(R_{n-1})}}~~,}\eqno{(3.1)}$$
where $S_D(R)$ is the surface area of a $D$-dimensional sphere of radius $R$:
$$S_D (R) = {{2\pi^{D/2}}\over{\Gamma (D/2)}} R^{D-1}~~.$$
Since the surface area $S_D(R)$ is proportional to $R^{D-1}$, the inward-walk
and outward-walk probabilities are
$$\eqalign{P_{\rm out}(n)&={{R^{D-1}_n}\over {R^{D-1}_n+R^{D-1}_{n-1}}}~~,\cr
P_{\rm
in}(n)&={{R^{D-1}_{n-1}}\over{R^{D-1}_n+R^{D-1}_{n-1}}}~~.}\eqno{(3.2)}$$
These formulas are valid for all $D$ and all $n\geq 2$. For $n=1$, $P_{\rm out}
(1)=1$, and $P_{\rm in}(1)=0$ because there is {\sl no} region inward from
$n=1$
and therefore the walker {\sl must} go outward. Note that $P_{\rm out}$ and
$P_{\rm in}$ are nonnegative numbers less than or equal to $1$ with
$P_{\rm in} (n)+P_{\rm out} (n)=1$; thus, they are
acceptable probabilities for all real $D$.$^7$

In this model, the probability $C_{n,t}$ that the random walker arrives at
$region n$ at
time~$t$ satisfies a recursion relation analogous to that in (2.2),
$$\eqalign{
C_{n,t}&=P_{\rm in}(n+1)C_{n+1,t-1}+P_{\rm out}(n-1)C_{n-1,t-1}\quad (n\geq
2)~~,\cr
C_{1,t} &= P_{\rm in}(2) C_{2,t}~~,}\eqno{(3.3)}$$
where $P_{\rm in}$ and $P_{\rm out}$ are given by (3.2). This equation states
that for a random walker to be in $region~n$ at time $t$, she must have come
from $region~n+1$ or $region~n-1$ at time $t-1$. The initial condition
satisfied
by $C_{n,t}$ is
$$C_{n,0} = \delta_{n,1}~~,\eqno{(3.4)}$$
which states that at time $t=0$ the random walker is located in the central
sphere ($region~1$).

We can now use (3.3) and (3.4) to calculate $C_{1,2}$, the probability of
being in the central sphere ($region~1$) at the second time step:
$$C_{1,2} = {{R^{D-1}_1}\over {R^{D-1}_2+R^{D-1}_1}}~~.$$
Since $R_2>R_1>0$ we conclude that $0 < C_{1,2} < 1$ and thus that $C_{1,2}$
[unlike $C_{{\bf 0},2}$ in (2.9)] is an acceptable probability for all real
values of $D$. For the special simple case of equally-spaced concentric spheres
for which $R_n = n$ we have plotted $C_{1,2}$, $C_{1,4}$, $C_{1,6}$ in Fig.~6.
Note that these probabilities do not oscillate like those in Figs.~1 and 2
and they are acceptable probabilities for all real $D$.

We present two methods for obtaining from (3.3) the value of $\Pi_D$,
the probability that a random walker who starts at the central sphere
eventually returns to the central sphere ({\sl region} 1). The first
technique involves converting the partial difference equation in
(3.3) to an ordinary difference equation by summing over $t$. We
begin by making the substitution
$$\eqalign{ C_{n,t} &= (R_n^{D-1} + R_{n-1}^{D-1})E_{n,t}\quad (n\geq 2)~~,\cr
C_{1,t} &= R_1^{D-1} E_{1,t}~~.}\eqno{(3.5)}$$
Then the equation satisfied by $E_{n,t}$ is
$$\eqalign{(R_n^{D-1}+R_{n-1}^{D-1})E_{n,t} &= R_n^{D-1} E_{n+1,t-1}
+ R_{n-1}^{D-1}E_{n-1,t-1}~~(n\geq 2)~~,\cr
E_{1,t} &= E_{2,t-1}~~,}\eqno{(3.6)}$$
where we have substituted the values of $P_{\rm in}$ and $P_{\rm out}$ in
(3.2).
In addition, the first line in (3.6) is also valid for $n=1$ and $D>1$ if we
define $R_0 = 0$.

We can eliminate the $t$ index from (3.6) by summing over $t$. We define
$$F_n = \sum_{t=0}^{\infty} E_{n,t}\quad (n\geq 1)\eqno{(3.7)}$$
and sum (3.6) from $t=1$ to $t=\infty$. [Of course, we must assume
here that the result is finite. If the sum over $C_{n,t}$ is infinite
then from (2.11) we have immediately that $\Pi_D=1$.] The result is
$$\eqalign{(R_n^{D-1}+R_{n-1}^{D-1})F_n &= R_n^{D-1} F_{n+1}
+ R_{n-1}^{D-1}F_{n-1}~~(n\geq 2)~~,\cr
R_1^{D-1} \left ( F_1 - F_2 \right ) &= 1~~,}\eqno{(3.8)}$$
where we have made use of the initial condition in (3.4).

The first line in (3.8) can be rewritten in exact-difference form:
$$[R_n^{D-1} F_n - R_{n-1}^{D-1} F_{n-1}] = [R_n^{D-1} F_{n+1}-R_{n-1}^{D-1}
F_n] ~(n\geq 2)~~,\eqno(3.9)$$
which we then sum from $n=2$ to $n=N-1$ to give
$$R_{N-1}^{D-1} F_{N-1} - R_1^{D-1} F_1 = R_{N-1}^{D-1} F_N -
R_1^{D-1} F_2~~(N\geq 3).\eqno(3.10)$$
We eliminate $F_2$ from (3.10) using the second line in (3.8) and obtain
$$ [F_N - F_{N-1}] = - R_{N-1}^{-(D-1)}~~(N\geq 2)\eqno(3.11)$$
and sum it from $N=2$ to $N=K$. The result is
$$F_K = F_1 - \sum_{N=1}^{K-1} {1\over{R_N^{D-1}}}~~.\eqno{(3.12)}$$

If $F_K\to 0$ as $K\to\infty$ (we will establish the validity of this
limit later) then we have the simple result that
$$\sum_{t=0}^{\infty} C_{1,t}=\sum_{N=1}^{\infty}\left ({{R_1}\over{R_N}}
\right ) ^{D-1}\eqno{(3.13)}$$
from which we can express the probability that a random walker returns to
the origin as a generalized zeta function
$$\Pi_D = 1- {1\over { \sum_{N=1}^{\infty} \left ({{R_1}\over{R_N}}
\right )^{D-1}  }}~~.\eqno{(3.14)}$$

More generally, for arbitrary functions $P_{\rm in} (n)$ and $P_{\rm out} (n)$
with $P_{\rm in} (n) + P_{\rm out} (n) = 1$, $n\geq 2$, and $P_{\rm in} (1) =
0$
and $P_{\rm out} (1) = 1$, we obtain from (3.3)
$$\sum_{t=0}^{\infty} C_{1,t}=1+\sum_{N=2}^{\infty}~\prod_{n=2}^N~
{{P_{\rm in} (n)}\over{P_{\rm out} (n)}}~~,\eqno{(3.15)}$$
and
$$\Pi_D=1-{1\over{1+\sum_{N=2}^{\infty}~\prod_{n=2}^N~
{{P_{\rm in} (n)}\over{P_{\rm out} (n)}}}}~~.\eqno{(3.16)}$$
The results in (3.15) and (3.16) reduce  to (3.13) and (3.14),
respectively, for the choice of $P_{\rm in} (n)$ and $P_{\rm out} (n)$
in (3.2).
\bigskip
\bigskip
\noindent
{\bf Alternative Solution in Terms of Continued Fractions}
\bigskip
If we simply iterate (3.3) we obtain a sequence of increasingly complicated
formulas for $C_{1,2t}$:
$$\eqalign{
C_{1,0}&= 1~~,\cr
C_{1,2}&= Q_1~~,\cr
C_{1,4}&=Q_1 (Q_1 + Q_2 )~~,\cr
C_{1,6}&= Q_1 [ (Q_1+Q_2)^2 + Q_2Q_3 ]~~,}
\eqno{(3.17)}$$
and so on, where
$$ Q_n = P_{\rm in} (n+1) P_{\rm out} (n)~~. \eqno{(3.18)}$$
However, the sequence of formulas in (3.17) is a clearly recognizable
pattern:$^8$
 $C_{1,2t}$ are the coefficients in the Taylor series for the function
whose continued-fraction representation has the coefficients $Q_n$:
$$\sum_{t=0}^{\infty} C_{1,2t} x^t = 1/(1-Q_1 x/(1 - Q_2 x/(1 - Q_3 x/(
1 - Q_4 x/ \ldots ))))~~.\eqno{(3.19)}$$
Thus,
$$\sum_{t=0}^{\infty} C_{1,2t} = 1/(1-Q_1/(1-Q_2/(1-Q_3/(1-Q_4/\ldots ))))
\eqno{(3.20)}$$
and
$$\Pi_D = Q_1/(1-Q_2/(1-Q_3/(1-Q_4/(1-Q_5/\ldots))))~~.\eqno{(3.21)}$$

{}From (3.20) it is easy to recover (3.15) for arbitrary $P_{\rm in}$
and $P_{\rm out}$, or (3.13) using $Q_1 = R_1^{D-1}/(R_1^{D-1}
+ R_2^{D-1})$ and $Q_n = R_n^{2D-2}/[(R_{n-1}^{D-1}+R_n^{D-1})(R_n^{D-1}+
R_{n+1}^{D-1})],~n\geq 2$, by the following procedure. For instance,
if we truncate (3.20) before the first minus
sign we obtain the first term in the sum on the right side of (3.13); if we
truncate before the second minus sign we obtain the first two terms in the sum;
and so on. Thus, as a byproduct of this analysis we have established the
continued-fraction identity
$$\eqalign{\sum_{n=1}^{\infty} {{S_1}\over{S_n}}&= 1/\left(1-{{S_1}\over{S_1 +
S_2}}/
\left(1-{{S_2^2}\over {(S_1+S_2)(S_2+S_3)}}/
\left(1-{{S_3^2}\over {(S_2+S_3)(S_3+S_4)}}/
\ldots \right)\right)\right)\cr
&= 1/\left(1-S_1/\left(S_1+S_2-S_2^2/\left(S_2+S_3 - S_3^2/
\left(S_3+S_4-S_4^2/
\ldots \right)\right)\right)\right)~~,} \eqno{(3.22)}$$
which contains as a special case a continued-fraction representation of the
zeta function:
$$\zeta (r) = 1/(1-1/(1+2^r-2^{2r}/(2^r+3^r-3^{2r}/(3^r+4^r-4^{2r}/\ldots))))~~
    .\eqno{(3.23)}$$

Observe that this method for establishing (3.14) avoids the necessity of
proving\break
$\lim_{K\to\infty} F_K = 0$ in (3.12) because the right side of
(3.12) vanishes explicitly in this limit.

\bigskip
\bigskip
\noindent{{\bf Special Case} ${\bf R}_{\bf n} {\bf =} {\bf n}$}
\bigskip

An interesting special and simple case of (3.14) arises if we set $R_n =
n$.$^9$
Note that $R_n=n$ corresponds to equally-spaced concentric spheres. For this
case
$$\Pi_D = 1-{1\over {\zeta (D-1)}}~~,\eqno{(3.24)}$$
where $\zeta$ is the Riemann zeta function. For this case we have established
that for $D>2$, $F_K$ vanishes as $K\to\infty$. Specifically, we find that for
$R_n = n$, $F_K$ vanishes like $K^{2-D}$ as $K\to\infty$.$^{10}$

The structure of (3.24) is reminiscent of an interesting probabilistic result
in
number theory. It is not difficult to show that the probability of a randomly
chosen integer being evenly divisible by (contains as a factor) the $N$th power
of some integer is $1-1/\zeta (N)$ for $N\geq 2$.

Although the solution to the partial difference equation in (3.3) is
complicated in
general for arbitrary $D$, when $R_n = n$ we have been able to obtain exact
solutions for the
special cases of integer dimension $D=0,~1$ and $2$. The simplest case is
$D=1$. For this case
$$\eqalign{
C_{n,n+2j-1} &={{(n+2j-1)!}\over { (n+j-1)! j! 2^{n+2j-2}}}\quad (n\geq
2)~~,\cr
C_{1,2t} &= {{(2t)!}\over {t!t!2^{2t}}}~~.}\eqno{(3.25)}$$
Using the generating function formula
$$\sum_{t=1}^{\infty} P_{2t} x^t = 1 - {1\over {\sum_{t=0}^{\infty}
C_{1,2t} x^t }}~~,\eqno{(3.26)}$$
which corresponds to (2.10), we find that
$$P_{2t}(D=1) = {{(2t)!}\over {t!t!(2t-1) 2^{2t}}}~~.\eqno{(3.27)}$$
Using (3.27) we verify directly that
$$\eqalign{
\Pi_1 &= \sum_{t=1}^{\infty} P_{2t} x^{2t-1} \Bigm |_{x=1}\cr
&= \int_{0}^1 {{dx}\over{x^2}} \sum_{t=1}^{\infty} {{\Gamma (t+1/2)
x^{2t}}\over {t!\sqrt{\pi}}}\cr
&= \int_{0}^1 {{dx}\over {x^2}} \left ( {1\over{\sqrt{1-x^2}}}-1\right )\cr
&=1~~.  }\eqno{(3.28)}$$

For the case $D=2$, we have
$$C_{n,n+2j-1}(D=2)={{(2n-1)(n+2j-1)!}\over {j!(2n+2j-1)!!2^j}}~~.
\eqno{(3.29)}$$
Thus,
$$C_{1,2t} = {1\over {1+2t}}~~.\eqno{(3.30)}$$
Using (3.26) we can obtain from (3.30) the probabilities $P_{2t}$ and we
find that they are not simple numbers:
$$\eqalign{
P_2 &= {{1} \over {3}}~~,\cr
P_4 &= {{4} \over {45}}~~,\cr
P_6 &= {{44} \over {945}}~~,\cr
P_8 &= {{428} \over {14175}}~~,\cr
P_{10} &= {{10196} \over {467775}}~~,\cr
P_{12} &= {{10719068} \over {638512875}}~~.}\eqno{(3.31)}$$
Nevertheless, we easily calculate $\Pi_2$:
$$\eqalign{
\Pi_2 &={x\over 3}+{{4x^2}\over{45}}+{{44x^3}\over{945}}+\ldots\Bigm |_{x=1}\cr
&=1-{1\over{1+{x\over 3}+{{x^2}\over 5}+{{x^3}\over 7}+{{x^4}\over 9}
+\ldots}} \Bigm |_{x=1}\cr
&= 1 - {1\over {1+{1\over y} \left (-y + {1\over 2} \ln
{{1+y}\over{1-y}}\right)
}}\Bigm |_{y=\sqrt{x}=1}\cr &=1~~.}\eqno{(3.32)}$$

The case $D=0$ is slightly more complicated. For this case there is no
simple formula for $C_{n,t}$. Rather, we must consider special cases:
For $n>1$ we have
$$\eqalign{
C_{n,n-1} &= B_{n,0}~~,\cr
C_{n,n+1} &= B_{n,1} + {2\over 3}B_{n,0}~~,\cr
C_{n,n+3} &= B_{n,2} + {2\over 3}B_{n,1} + {{26}\over {45}}B_{n,0}~~,\cr
C_{n,n+5} &= B_{n,3} + {2\over 3} B_{n,2} + {{26}\over{45}} B_{n,1} +
{{502}\over{945}} B_{n,0}~~,}\eqno{(3.33)}$$
and so on, where
$$B_{n,j} = {{(n+2j-2)! (2n-1)}\over {j! (2n+2j-1)!! 2^j}}~~.$$
The numbers $C_{n,t}$ corresponding to $n=1$ are
$$\eqalign{
C_{1,0} &= 1~~,\cr
C_{1,2} &= {{2} \over {3}}~~,\cr
C_{1,4} &= {{26} \over {45}}~~,\cr
C_{1,6} &= {{502} \over {945}}~~,\cr
C_{1,8} &= {{7102} \over {14175}}~~,\cr
C_{1,10} &= {{44834} \over {93555}}~~.}\eqno{(3.34)}$$
In contrast, the formula for $P_{2t}$ is extremely simple:
$$P_{2t} = {2\over {(2t+1)(2t-1)}}~~.\eqno{(3.35)}$$
Hence,
$$\Pi_0 =  \sum_{t=1}^{\infty} {2\over {(2t+1)(2t-1)}}=1~~.\eqno{(3.36)}$$

\bigskip
\bigskip
\leftline{\bf Reciprocity Property of (3.3)}
\bigskip
Finally, we mention an interesting reciprocity property that is exhibited by
the
probabilities $C_{1,2t}$ and $P_{2t}$. This property is associated with
reflecting the dimension $D$ about the point $D=1$:
$$ \left [P_{2t}\right ] \Bigm |_{D} =
\left [C_{1,2t-2} - C_{1,2t}\right ] \Bigm |_{2-D}\quad(t\geq
1)~~.\eqno{(3.37)}$$
An immediate consequence of this probability can be obtained by summing
(3.37) from $t=1$ to $\infty$:
$$\eqalign{
\Pi_D &= \sum_{t=1}^{\infty} \left [P_{2t}\right ] \Bigm |_{D}\cr
&=\left [C_{1,0} - C_{1,\infty}\right ] \Bigm |_{2-D}\cr
&= 1- C_{1,\infty} \Bigm |_{2-D} ~~.}\eqno{(3.38)}$$
Hence $\Pi_D = 1$ if and only if $\lim_{t\to\infty} C_{1,2t} = 0$ in
$(2-D)$ dimensions.

We can prove the reciprocity relation in (3.37) in completely general
terms. Let $C_{n,t}$ be the solution of (3.3-4), and let ${\tilde C}_{n,t}$
be the solution of (3.3-4) with $P_{\rm in}(n)$ and $P_{\rm out}(n)$
{\sl interchanged} for each $n\geq 2$ but not $n=1$. Then,
$\sum_{t=0}^{\infty} C_{1,2t} x^t$ is given by (3.19) with $Q_n$ defined
in (3.18), and $\sum_{t=0}^{\infty} {\tilde C}_{1,2t} x^t$ is given by
(3.19) with $Q_n$ replaced by
$$\eqalign{{\tilde Q}_n&=P_{\rm out}(n+1) P_{\rm in}(n)\quad(n\geq2)~~,\cr
{\tilde Q}_1&=P_{\rm out}(2)~~.}\eqno(3.39)$$
With these definitions we obtain according to an identity given by Wall:$^{11}$
$$\eqalign{{1 \over {1-x}}=&1/(1-Q_1x/(1-Q_2x/(1-Q_3x/(1-Q_4x/\ldots))))\cr
&\times 1/(1-{\tilde Q}_1 x/(1 - {\tilde Q}_2 x/(1 - {\tilde Q}_3 x/(1 -
{\tilde Q}_4 x/ \ldots ))))~~.
}\eqno{(3.40)}$$
{}From (3.19) we then get
$${1\over {1-x}}=\left(\sum_{t=0}^{\infty} C_{1,2t} x^t \right)
\left(\sum_{t=0}^{\infty} {\tilde C}_{1,2t} x^t\right)~~.\eqno{(3.41)}$$
Using the generating function in (3.26) we obtain
$$\sum_{t=1}^{\infty} P_{2t} x^t - 1=\left(x-1\right)\sum_{t=0}^{\infty}
{\tilde C}_{1,2t} x^t ~~;\eqno{(3.42)}$$
the coefficient of $x^t$ in this equation is
$$P_{2t}={\tilde C}_{1,2t-2}-{\tilde C}_{1,2t} \quad(t\geq 1)~~.\eqno(3.43)$$
For the particular case in (3.2), interchanging $P_{\rm in}(n)$ and
$P_{\rm out}(n)$ for $n\geq 2$ corresponds to replacing $D$ by $(2-D)$
and relation (3.37) follows immediately.

\bigskip
\bigskip
We thank M. Moshe for useful discussions. LRM thanks the Physics Department at
Washington University for its hospitality.  We thank the U. S. Department of
Energy
for funding this research and one of us, SB, thanks the U. S. Department of
Education
for financial support in the form of a National Need Fellowship.
\vfill \eject

\leftline{\bf APPENDIX}
\bigskip

In Sec.~II we found an expression in (2.12) for the probability $\Pi_D$ of
a random walker eventually returning to the origin on a hypercubic lattice in
arbitrary integer dimension~$D$. For real values of $D>2$, (2.12) describes an
acceptable analytic continuation off the integers for such a probability. But
at
$D=2$, (2.12) approaches the correct result $\Pi_2=1$ from above with a nonzero
slope of $-\pi/2$. Thus, the physical probability must have a discontinuous
derivative as a function of $D$ at $D=2$. Here we calculate an asymptotic
series
for the function that analytically continues (2.12) smoothly to its nonphysical
values for $D<2$. The results indicate that this function has a singularity at
$D=1$. Furthermore, this asymptotic series provides excellent numerical results
for all $D>1$ which can be systematically improved at least for integer values
of $D$.

We start by rewriting the integral representation of $C_{{\bf 0},2t}$ in
(2.7) (recall that $C_{{\bf 0},t}=0$ when $t$ is odd):
$$C_{{\bf 0},2t}~=~\int_{0}^{\pi} {{d^Dk}\over{\pi^D}}~\exp\biggl[2t
\ln\left({1\over D} \sum_{j=1}^D \cos k_j \right) \biggr]~~.\eqno(A.1)$$
Now we consider $C_{{\bf 0},2t}$ in the limit of a large number of time steps,
$t\to\infty$. In this limit we can evaluate the integral in (A.1) by Laplace's
method$^{12}$. The exponent in the integral has a maximum either when all $D$
components of ${\bf k}$ are small or when all $D$ components of ${\bf k}$ are
close to $\pi$. Both of these regions give
identical contributions to the integral, and we can write
$$\eqalign{C_{{\bf 0},2t}~&\sim~2\int_{0}^{\epsilon}
{{d^Dk}\over{\pi^D}}~\exp\biggl[2t \ln\left(1-{1\over 2D}
\sum_{j=1}^D k_j^2+{1\over 24D}\sum_{j=1}^D k_j^4-\ldots \right)
\biggr] \cr
&~~~~~~~(t^{-1/2}\ll\epsilon\ll t^{-1/4}~~,~~~~t\to\infty )~~.}\eqno(A.2)$$
We scale each of the integration variables in (A.2) by $\sqrt{t/D}$ and
collect terms order by order in powers of $1/t$. Note that the choice of
$\epsilon$ in the end allows one to extend the integration region out
to infinity to within exponentially small corrections:
$$\eqalign{C_{{\bf 0},2t} &\sim 2\left ({D\over t}\right ) ^{D/2}
\int_{0}^{\epsilon\sqrt{t}} {{d^Dk}\over{\pi^D}} ~\exp\biggl[-
\sum_{j=1}^D k_j^2+{D\over 12t}\sum_{j=1}^D k_j^4-
{1\over{4t}}\left(\sum_{j=1}^D k_j^2\right)^2 + \ldots \biggr]~~  \cr
\noalign{\smallskip}
&~~~~~~~(1\ll\epsilon\sqrt{t}\ll t^{1/4},~t\to\infty)~~,\cr
\noalign{\bigskip}
C_{{\bf 0},2t} &\sim 2\left({D\over{4\pi t}}\right)^{D/2}
\int_{0}^{\infty} \left(\prod_{j=1}^D {{dk_j~e^{-
k_j^2}}\over{\sqrt{\pi}/2}}\right)~\biggl\{1 + {1\over
t}\left [ {D\over 12}\sum_{j=1}^D k_j^4-{1\over{4}}\left(\sum_{j=1}^D
k_j^2\right)^2\right ] + \ldots \biggr\}\cr
\noalign{\smallskip}
&~~~~~~~(t\to\infty)~~.}\eqno(A.3)$$
The remaining integrals separate into products of ordinary integrals
to any order in powers of $1/t$ and can be solved with simple
combinatorical considerations. We finally obtain
$$C_{{\bf 0},2t}~\sim~2\left({D\over{4\pi}}\right)^{D/2}
\sum_{n=0}^{\infty}~w_n(D)~t^{-n-D/2}\quad (t\to\infty)~~,\eqno(A.4)$$
where the $w_n(D)$ are polynomials in $D$ having rational coefficients:

$$w_0(D)~=~1~~,$$

$$w_1(D)~=~-{1\over 8}D~~,$$

$$w_{2}(D)~=~{{1}\over{96}}D-{{1}\over{128}}D^{2}
+{{1}\over{192}}D^{3}~~,$$

$$w_{3}(D)~=~{{3}\over{256}}D^{2}-{{9}\over{1,024}}D^{3}
+{{1}\over{512}}D^{4}~~,$$

$$w_{4}(D)~=~-{{1}\over{3,840}}D-{{143}\over{18,432}}D^{2}
+{{179}\over{12,288}}D^{3}-{{2,621}\over{294,912}}D^{4}
+{{69}\over{40,960}}D^{5}+{{1}\over{73,728}}D^{6}~~,$$

$$w_{5}(D)~=~{{25}\over{6,144}}D^{2}-{{3,005}\over{147,456}}D^{3}
+{{2,389}\over{98,304}}D^{4}-{{26,783}\over{2,359,296}}D^{5}
+{{359}\over{196,608}}D^{6}+{{7}\over{589,824}}D^{7}~~,$$

$$\eqalign{w_{6}(D)~&=~{{1}\over{32,256}}D-{{721}\over{368,640}}D^{2}
+{{656,987}\over{26,542,080}}D^{3}-{{632,981}\over{11,796,480}}D^{4}
+{{1,305,539}\over{28,311,552}}D^{5}\cr
&~~~-{{3,307,553}\over{188,743,680}}D^{6}
+{{4,828,249}\over{1,981,808,640}}D^{7}+{{193}\over{15,728,640}}D^{8}
+{{1}\over{42,467,328}}D^{9}~~,}$$

$$\eqalign{w_{7}(D)~&=~{{35}\over{36,864}}D^{2}-{{16,249}\over{589,824}}D^{3}
+{{4,432,691}\over{42,467,328}}D^{4}-{{4,639,363}\over{31,457,280}}D^{5}\cr
&~~~+{{22,302,679}\over{226,492,416}}D^{6}
-{{3,166,907}\over{100,663,296}}D^{7}
+{{1,743,949}\over{452,984,832}}D^{8}\cr
&~~~+{{5,491}\over{377,487,360}}D^{9}+{{11}\over{339,738,624}}D^{10}~~,}$$

$$\eqalign{w_{8}(D)~&=~-{{1}\over{122,880}}D
-{{302,179}\over{619,315,200}}D^{2}
+{{14,453,681}\over{495,452,160}}D^{3}
-{{13,334,609,269}\over{71,345,111,040}}D^{4}\cr
&~~~+{{696,626,939}\over{1,698,693,120}}D^{5}
-{{174,592,096,943}\over{407,686,348,800}}D^{6}
+{{4,230,152,963}\over{18,119,393,280}}D^{7}\cr
&~~~-{{391,310,431,607}\over{6,088,116,142,080}}D^{8}
+{{356,173,841}\over{50,734,301,184}}D^{9}
+{{226,174,807}\over{11,415,217,766,400}}D^{10}\cr
&~~~+{{1,111}\over{27,179,089,920}}D^{11}
+{{1}\over{32,614,907,904}}D^{12} ~,}$$

$$\eqalign{w_{9}(D)~&=~{{83}\over{327,680}}D^{2}
-{{49,679,143}\over{1,651,507,200}}D^{3}
+{{418,240,717}\over{1,321,205,760}}D^{4}
-{{198,111,107,857}\over{190,253,629,440}}D^{5}\cr
&~~~ +{{4,345,476,319}\over{2,717,908,992}}D^{6}
-{{1,437,041,954,251}\over{1,087,163,596,800}}D^{7}
+{{17,640,297,037}\over{28,991,029,248}}D^{8}\cr
&~~~-{{88,461,257,153}\over{601,295,421,440}}D^{9}
+{{29,436,108,509}\over{2,029,372,047,360}}D^{10}
+{{938,914,279}\over{30,440,580,710,400}}D^{11}\cr
&~~~+{{11,683}\over{217,432,719,360}}D^{12}
+{{5}\over{86,973,087,744}}D^{13}~,}$$

$$\eqalign{w_{10}(D)~&=~{{1}\over{270,336}}D-{{30,263}\over{247,726,080}}D^{2}
+{{1,819,137,181}\over{59,454,259,200}}D^{3}
-{{46,045,222,579}\over{89,181,388,800}}D^{4} \cr
&~~~+{{85,060,789,019,063}\over{34,245,653,299,200}}D^{5}
-{{9,858,288,912,307}\over{1,826,434,842,624}}D^{6}\cr
&~~~+{{1,740,371,777,264,779}\over{273,965,226,393,600}}D^{7}
-{{176,289,185,165,713}\over{40,587,440,947,200}}D^{8}\cr
&~~~+{{15,166,408,871,756,519}\over{8,766,887,244,595,200}}D^{9}
-{{871,621,884,030,119}\over{2,337,836,598,558,720}}D^{10}\cr
&~~~+{{6,459,557,472,735,757}\over{192,871,519,381,094,400}}D^{11}
+{{79,597,124,879}\over{1,461,147,874,099,200}}D^{12}\cr
&~~~+{{167,726,257}\over{2,191,721,811,148,800}}D^{13}
+{{601}\over{6,957,847,019,520}}D^{14}\cr
&~~~+{{1}\over{31,310,311,587,840}}D^{15}~~,}$$

$$\eqalign{w_{11}(D)~&=~{{11}\over{196,608}}D^{2}
-{{61,231,223}\over{1,981,808,640}}D^{3}
+{{390,360,741,901}\over{475,634,073,600}}D^{4} \cr
&~~~-{{4,043,100,547,039}\over{713,451,110,400}}D^{5}
+{{930,539,498,230,171}\over{54,793,045,278,720}}D^{6}\cr
&~~~-{{19,033,279,103,719}\over{695,784,701,952}}D^{7}
+{{57,117,406,563,298,579}\over{2,191,721,811,148,800}}D^{8}\cr
&~~~-{{44,382,762,511,849,337}\over{2,922,295,748,198,400}}D^{9}
+{{74,780,092,063,706,731}\over{14,027,019,591,352,320}}D^{10}\cr
&~~~-{{3,876,663,970,072,123}\over{3,740,538,557,693,952}}D^{11}
+{{11,991,012,720,744,317}\over{140,270,195,913,523,200}}D^{12}\cr
&~~~+{{60,295,396,519}\over{556,627,761,561,600}}D^{13}
+{{420,916,787}\over{3,506,754,897,838,080}}D^{14}\cr
&~~~+{{21,011}\over{166,988,328,468,480}}D^{15}
+{{19}\over{250,482,492,702,720}}D^{16}~~,}$$

$$\eqalign{w_{12}(D)~&=~-{{691}\over{268,369,920}}D
-{{69,760,249}\over{2,288,988,979,200}}D^{2}
+{{38,695,325,513}\over{1,245,708,288,000}}D^{3}\cr
&~~~-{{482,851,775,889,419}\over{376,702,186,291,200}}D^{4}
+{{3,429,357,841,822,897}\over{273,965,226,393,600}}D^{5}\cr
&~~~-{{1,001,400,307,492,157,137}\over{19,725,496,300,339,200}}D^{6}
+{{2,388,031,903,466,391,071}\over{21,917,218,111,488,000}}D^{7}\cr
&~~~-{{20,467,165,251,552,205,481}\over{147,283,705,709,199,360}}D^{8}
+{{7,793,853,343,774,704,271}\over{70,135,097,956,761,600}}D^{9} \cr
&~~~-{{94,989,731,986,215,946,793}\over{1,683,242,350,962,278,400}}D^{10}
+{{99,366,320,397,537,803,389}\over{5,610,807,836,540,928,000}}D^{11} \cr
&~~~-{{2,787,056,465,380,319,491,493}\over{888,751,961,308,082,995,200}}D^{12}
+{{10,961,446,588,181,276,321}\over{45,848,315,464,305,868,800}}D^{13} \cr
&~~~+{{24,950,506,464,410,431}\over{103,687,728,819,276,349,440}}D^{14}
+{{292,872,847,633}\over{1,402,701,959,135,232,000}}D^{15} \cr
&~~~+{{318,805,229}\over{1,683,242,350,962,278,400}}D^{16}
+{{59}\over{445,302,209,249,280}}D^{17} \cr
&~~~+{{1}\over{36,069,478,949,191,680}}D^{18}~~,}\eqno(A.5)$$
and so on. The polynomials $w_n(D)$ have a few remarkable properties.
For instance, the degree of the $n$th polynomial is $[3n/2]$. The
coefficient of $D$ in the $n$th polynomial, $n\geq 1$, is $B_n 2^{-n-1}/n$
where the $B_n$ are the Bernoulli numbers. Also, the coefficient of
$D^{3n/2}$ in the $2n$th polynomial is $192^{-n}/n!$ and the coefficient of
$D^{1+3n/2}$ in the $2n+1$st polynomial is ${{4n-1}\over {8n!}} 192^{-n}$.

To obtain an approximation for $\Pi_D$ in (2.12) we sum (A.4) over $t$.
Using the definition of a generalized Riemann $\zeta$-function,
$$\zeta(s,q)~=~\sum_{n=q}^{\infty} ~n^{-s}~~,\eqno(A.6)$$
we define for $N\geq0$:
$$H_D^N(x)~=~2\left({D\over{4\pi}}\right)^{D/2} \sum_{n=0}^{\infty}~
w_n(D) ~\zeta(n+D/2,N+1)~ x^n~~.\eqno(A.7)$$
Then we expect that
$$\Pi_D~\sim~1~-~{1\over{\sum_{t=0}^N C_{{\bf 0},2t}+H_D^N(1)}}\quad
(N\to\infty)~~.\eqno(A.8)$$
We obtain excellent numerical results from (A.8) by the following
procedure based on ``optimal truncation'':$^{13}$ For given $D$ and fixed
$N$ truncate the series for $H_D^N(x)$ in (A.7) right {\sl before}
the term with the smallest coefficient of $x^n$ and evaluate the
truncated series at $x=1$. Inserting this value for $H_D^N(1)$ in
(A.8) provides a very accurate numerical approximation for $\Pi_D$.
The relative size of the first omitted coefficient in the series
in (A.7) usually gives a good estimate of the relative error for
the value of $\Pi_D$. Fig.~3 contains the plot of $\Pi_D$ in (A.8) for $N=3$
as a function of $D$ for $1<D<2$. The choice of $N=3$ ensures that the
error in this plot is well below the width of the line. Note the
smooth continuation at $D=2$ with the plot of $\Pi_D$ in (2.12).

The function in (A.8) appears to have a singularity at $D=1$: For increasing
$N$, the sum of the two terms $\sum_{t=0}^N C_{{\bf 0},2t}$ and $H_1^N(1)$ in
(A.8) appears to approach zero. This cancellation is nontrivial because at
$D=1$
the former is a rational number while the latter is a sum over
$\zeta$-functions
of half-integer arguments. For $N=1$, the cancellation is already accurate to
$0.002\%$. For $N=15$, for example, we find that
$$\sum_{t=0}^{15} C_{{\bf 0},2t}=300,540,195/67,108,864$$
and when this rational number is rewritten in decimal form the first $16$
digits
are canceled by $H_1^{15}(1)$. A numerical study indicates that the singularity
is a simple pole with a residue close to $2$.

For integer values of $D$, this approximation can be improved systematically by
increasing $N$. As an example, we have calculated the value of $\Pi_3$ in
(2.14)
to $0.000003\%$ accuracy for $N=4$. In Table 2 we list the values thus obtained
for $\Pi_3$ for $N=1,~2,~3$ and $4$ and the order at which the truncation
occurs. Note that even $N=1$ gives an accuracy of $0.01\%$ in this case. In
Table 3 we list the values of $\Pi_D$ for the integer dimensions from $D=3$ to
$10$ that were obtained up to the indicated accuracy with this method.

For noninteger values of $D$, increasing $N$ generally does not improve the
numerical accuracy for $\Pi_D$ in (A.8). For a given noninteger value of $D$,
$N$ eventually becomes so large that $C_{{\bf 0},2N}$ ceases to be a
probability, as we pointed out in Sec.~II. Apparently, the asymptotic series in
(A.7) cannot correct for this effect and the values for $\Pi_D$ finally
diverge.
\vfill \eject

\centerline{\bf REFERENCES}
\bigskip
\item{$^1$} $D$-dimensional quantum-field-theory calculations are first
introduced in C. M. Bender, S. Boettcher, and L. Lipatov,
Phys.~Rev.~Lett.~{\bf 68}, 3674 (1992).
\medskip
\item{$^2$} A more detailed treatment of $D$-dimensional quantum field theory
methods may be found in C. M. Bender, S. Boettcher, and L. Lipatov,
Phys.~Rev.~D
{\bf 46}, 5557 (1992).
\medskip
\item{$^3$} New results on the Ising limit of quantum field theory have been
obtained using dimensional expansions; see C. M. Bender and S. Boettcher,
Phys.~Rev.~D (to be published).
\medskip
\item{$^4$} The geometry that we have devised may be useful for performing
field-theoretic calculations in arbitrary noninteger dimension.
\medskip
\item{$^5$} Essentials of random walks may be found in C.~Itzykson and
J.-M.~Drouffe, {\sl Statistical Field Theory} (Cambridge University Press,
Cambridge, 1989), Vol. 1. Note that the result in (2.14) is incorrect in this
and the reference quoted therein.
\medskip
\item{$^6$}
Spherical symmetry allows us to assume in this motivational discussion that the
walker is located at any point in the volume labeled $n$ with equal likelihood.
This is of course not true for the hypercubic lattice. For example, when $t=2$
and $D=2$ (see Fig.~4), the random walker may reach a square labeled $3$ two
units to the right of $1$ or else one unit to the right and one unit above $1$.
In the former case the probability of reaching such a square is  $1/4 \times
1/4
=1/16$ while in the latter case the probability of reaching such a square is
$1/4\times 1/2 = 1/8$. This difference reflects the fact that the hypercubic
lattice is not spherically symmetric. In the spherical geometry pictured in
Fig.~5 the random walker is equally likely to be at any point in the volume
between the spheres.
\medskip
\item{$^7$}
Spherical surface areas in this model of random walks determine the
relative probability of going outward or inward.  However, this is not the only
way
to construct a model of a random walk. For an alternative random walk we could
generalize the notion of Buffon's needle: Suppose we are in $region~n$.
We randomly drop a needle whose length equals $R_n - R_{n-1}$, the radial
extent
of the region, in such a way that one end of the line segment lies in
$region~n$. Then, the random walker enters the region in which the other
end of the line segment lies. For very large concentric circles (where the
radius is large compared with the spacing) the circumference is locally flat
and
the probability of moving outward, inward, or remaining in the same region
approaches the usual Buffon's needle result that is discussed in
elementary calculus textbooks such as G. Klambauer, {\sl Aspects of
Calculus} (Springer, New York, 1986), p. 365.
\medskip
\item{$^8$}
Polynomial structures like that in (3.17) can be used to obtain
interesting and useful nonlinear transforms. See C. M. Bender and K.
A. Milton, J. Math. Phys. (to be published).
\medskip
\item{$^9$}
Here, we regard this case as special because in some sense it represents a
$D$-dimensional random walk in {\sl flat} space. We can think of curved space
as
having the property that the surface area of the consecutive nested spheres
does
not grow like $n^{D-1}$.
\medskip
\item{$^{10}$}
Entries in a table of values for the probabilities $C_{n,t}$ form an infinite
triangle on an $(n,t)$ grid. For a given value of $n\geq 1$, $n-1$ is
the smallest value of $t$ for which $C_{n,t}$ is nonzero. (This is
because the random walker cannot reach the $n$th
region until time $n-1$.) Thus, the lowest value of $t$ contributing to the sum
in (3.7) is $n-1$. If we examine this sum numerically we find that the terms in
the sum at first increase with $t$ and reach a maximum and then die off to $0$
as $t$ continues to increase. Also, for fixed $t$ as we vary $n$, we again
observe a local maximum. Thus, in the table of probabilities there is a
crest; this crest lies on a curve that has the shape of a
parabola: $t(D-1) = n^2$. This parabola is just the curve that corresponds to
similarity solutions of the continuous version (the parabolic partial
differential equation) of (3.3).
\medskip
\item{$^{11}$}
Relation (3.40) can be found in H. S. Wall, {\sl Analytic Theory of
Continued Fractions} (Van Nostrand, New York, 1948), p. 281, where
we identify $g_{n-1}=P_{\rm in}(n),~n\geq 2$.
\medskip
\item{$^{12}$}
Elementary presentation of Laplace's method may be found in C.~M.~Bender and
S.~A.~Orszag, {\sl Advanced Mathematical Methods for Scientists and Engineers}
(McGraw-Hill, New York, 1978), Chap.~6.
\medskip
\item{$^{13}$}
See Ref.~12, Chap.~3, for a discussion of optimal truncation of asymptotic
series.
\vfill \eject

\centerline{\bf FIGURE CAPTIONS}
\bigskip

\noindent
Figure 1. The first three nonzero probabilities $C_{{\bf 0},t}$ of returning
to the origin after $t=2,~4,~{\rm and}~6$ time steps, plotted as functions
of $D$. Observe that $C_{{\bf 0},t}$ is not an acceptable probability until
$D$ is sufficiently large.
\medskip
\noindent
Figure 2. Same as in Fig.~1 except that we have plotted $C_{{\bf 0},18}$
and $C_{{\bf 0},20}$. For small $D$ these functions become so large that it
is necessary to scale the vertical axis using the transformation $y \to
[2y/(1+|y|)]$. That is, $C = \pm\infty$ corresponds to $\pm 2$ on the vertical
axis while the interval between $0$ and $1$, where $C$ is a probability,
is mapped onto itself.
\medskip
\noindent
Figure 3. A plot of $\Pi_D$ in (2.12) representing the probability that a
random walker on a $D$-dimensional hypercubic lattice will eventually return
to the origin. The curve passes through $1$ at $D=2$ with a slope of
$-\pi/2$ and for large $D$ it falls off like $1/(2D)$. For $0\leq D<2$,
$\Pi_D \equiv 1$. However, if we analytically continue the function in
(2.12) to values of $D$ less than $2$ we obtain a function that grows as $D$
decreases and eventually blows up at $D=1$ (see Appendix).
\medskip
\noindent
Figure 4. Dual hypercubic lattice for random walks in two-dimensional space.
A random walker begins in the central square labeled $1$. On her first step she
enters one of the four squares labeled $2$. On her next step the walker either
returns to square $1$ or enters one of the eight squares labeled $3$. In
general, at each step the walker goes to an adjacent region, either inside or
outside her current region.  This way of visualizing a
random walk is completely equivalent to the conventional way in which the
walker goes from vertex to vertex on a hypercubic lattice.
\medskip
\noindent
Figure 5. The spherical geometry for the arbitrary-dimensional random walks
discussed in this paper. The random walker begins in the central region and
on her first step enters $region~2$. On her second step she either returns to
$region~1$ or moves outward to $region~3$ with a relative probability equal
to the surface area between her current region and the adjacent region to
which she goes.
\medskip
\noindent
Figure 6. The first three nonzero probabilities $C_{1,t}$ of returning
to the origin after $t=2,~4,~{\rm and}~6$ time steps, plotted as functions
of $D$, for a random walk in a geometry of equally-spaced concentric spheres.
Observe that in contrast with Figs.~1 and 2, $C_{1,t}$ {\sl is} an acceptable
probability for all values of $D$.
\medskip
\noindent
Figure 7. A plot of $\Pi_D$ in (3.24) representing the probability that a
random
walker on a $D$-dimensional concentric spherical geometry with equally-spaced
spheres will eventually return to the origin. Note that the curve is similar to
that in Fig.~3; it passes through $1$ at $D=2$ with a slope of $-1$ and for
large $D$ it falls off like $2^{1-D}$. For $0\leq D<2$, $\Pi_D \equiv 1$. If we
analytically continue the function in (3.24) to values of $D$ less than $2$ we
obtain a function that grows as $D$ decreases but remains finite until $D=-1$.
At $D=1$ it passes through $3$ and at $D=0$ it passes through $13$. There are
simple poles for all negative odd-integer values of $D$.
\vfill\eject

\centerline{\bf Table 1}
\bigskip
\noindent
Table 1. The critical value of the dimension, $D_{\rm critical}$, for which
$C_{{\bf 0},t}$ can first be considered a probability; that is, for all $D$
larger than $D_{\rm critical}$, $C_{{\bf 0},t}$ lies between $0$ and $1$.
\bigskip
\bigskip

$$\vbox{\halign{
\hfil#\hfil&\quad\hfil#\hfil\cr
$t$ & $D_{\rm critical}$ \cr
\noalign{\medskip\hrule\medskip}
2    &    0.500   \cr
6    &    0.722   \cr
10   &    1.285   \cr
14   &    1.795   \cr
18   &    1.990   \cr
22   &    2.444   \cr
26   &    3.488   \cr
30   &    3.892   \cr
34   &    3.997   \cr
38   &    4.000   \cr
42   &    5.647   \cr
46   &    5.955   \cr
50   &    6.000   \cr
54   &    7.243   \cr
58   &    7.777   \cr
}}$$
\vfill\eject

\centerline{\bf Table 2}
\bigskip
\noindent
Table 2. Convergence of the ``optimal truncation'' procedure to calculate
$\Pi_D$ for $D=3$. The exact result, $0.34053732\ldots$, is given in (2.14).
\bigskip
\bigskip
$$\vbox{\halign{\hfil#\hfil&\quad\hfil#\hfil&\quad\hfil#\hfil\cr
&&\hfill truncated\cr
$N$&$\Pi_3$&after order\cr
\noalign{\medskip \hrule \medskip}
1&0.340492587&4\cr
2&0.340536478&7\cr
3&0.340537202&9\cr
4&0.340537315&11\cr}}$$

\vfill\eject

\centerline{\bf Table 3}
\bigskip
\noindent
Table 3. Values of $\Pi_D$ in (2.12) for some integer values of $D$.
\bigskip
\bigskip
$$\vbox{\halign{\hfil#\hfil&\quad\quad#\hfil\cr
$D$&      $\Pi_D$\cr
\noalign{\medskip \hrule \medskip}
                 3&0.340537329551\cr
                 4&0.19320167322 \cr
                 5&0.135178609   \cr
                 6&0.104715496   \cr
                 7&0.085844934   \cr
                 8&0.07291265    \cr
                 9&0.06344775    \cr
                10&0.05619754    \cr}}$$
\vfill \eject
\bye